\documentstyle[preprint,prl,aps,epsf]{revtex}
\begin{document}
\title{Fourth Order Diffusion Monte Carlo Algorithms\\ 
for Solving Quantum Many-Body Problems
}
\author{Harald A. Forbert and Siu A. Chin}
\address{Center for Theoretical Physics, Department of Physics, 
Texas A\&M University, 
College Station, TX 77843}
\date{\today}
\maketitle
\begin{abstract}

By decomposing the important sampled imaginary time Schr\"odinger 
evolution operator to fourth order with positive coefficients, 
we derived a number of distinct fourth order
Diffusion Monte Carlo algorithms. These sophisticated 
algorithms require higher derivatives of the drift velocity and 
local energy and are more complicated to program. 
However, they allowed very large time steps to be used, 
converged faster with lesser
correlations, and virtually eliminated the step size error.
We demonstrated the effectiveness of these quartic algorithms by
solving for the ground state properties of bulk liquid helium.

\end{abstract}
\pacs{PACS:02.70.Lq,05.30.Jp,67.40.Db}
\section {Introduction}

The basic idea of the Diffusion Monte Carlo (DMC) algorithm is to solve for
the ground state of the Hamiltonian $H$ by evolving the imaginary time
Schr\"odinger equation
\begin{equation}
- {\partial \over \partial t } \psi ( {\bf x}, t ) =
H \psi ( {\bf x}, t )=\Bigr [-{1\over 2}\nabla^2+V({\bf x})\Bigl ]
\psi({\bf x},t)
\label{imsch}
\end{equation}
to large time\cite{anderson75,reynolds82,moskowitz82}. 
Here, ${\bf x}$ and $\nabla^2$ denote the 
coordinate and the Laplacian of the $N$-particle system.    
In order for the algorithm to be practical, capable of handling rapidly
varying potentials, it is essential to implement important sampling 
as suggested by Kalos\cite{kalos74}. This means
that instead of solving for $\psi ({\bf x})$, one evolves the product 
wave function $\rho ({\bf x})=\phi ({\bf x})\psi ({\bf x})$ 
according to\cite{reynolds82,moskowitz82}
\begin{eqnarray}
-{\partial \over \partial t} \rho({\bf x},t) &&= 
\phi({\bf x}) H \phi^{-1}({\bf x}) \rho({\bf x},t)
, \nonumber \\ 
&&=
-{1\over 2}\nabla^2\rho({\bf x},t)
+\nabla_i\Bigl [G_i({\bf x})\rho({\bf x},t)\Bigr ]
 +E_L({\bf x}) \rho({\bf x},t),\label{impeq} \\
 &&= \Bigr [T + D + E_L] \rho({\bf x},t)=\tilde H\rho({\bf x},t),
\label{impop}
\end{eqnarray}
where 
\begin{equation}
E_L({\bf x}) = \phi({\bf x})^{-1} H \phi({\bf x}),
\end{equation}
is the {\it local energy} 
\begin{equation}
G_i({\bf x})  = \phi({\bf x})^{-1}\nabla_i\phi({\bf x})=-\nabla_i S({\bf x}),
\end{equation}
is the {\it drift velocity} and 
$\phi ({\bf x})=\exp[-S({\bf x})]$ is the trial ground state wave function.

Eq.(\ref{impop}) has the formal operator solution
\begin{equation}
\rho (t) ={\rm e}^{-t(T+D+E_L )}\rho (0)
=\Bigl [{\rm e}^{-\epsilon(T+D+E_L )}\Bigr ]^n\rho (0).
\label{eop1}
\end{equation}
Various DMC algorithms correspond to different approximations
to the short time evolution operator ${\rm e}^{-\epsilon(T+D+E_L )}$. 
Initial implementations\cite{anderson75,reynolds82,moskowitz82}
of the DMC algorithm correspond to essentially
approximating
\begin{equation}
{\rm e}^{-\epsilon(T+D+E_L )}\approx 
{\rm e}^{-{1\over 2}\epsilon E_L }
{\rm e}^{-\epsilon T }
{\rm e}^{-\epsilon D }
{\rm e}^{-{1\over 2}\epsilon E_L },
\end{equation}
which is at most first order in $\epsilon$. By use of various clever 
tricks, this error can be reduced substantially in specific
applications\cite{umrigar93}. 
However, it was recognized by Chin\cite{chin90}
that in order to have a general second order DMC algorithm, one must simulate the 
embedded Fokker-Planck evolution operator ${\rm e}^{-\epsilon(T+D)}$,
{\it i.e.} the Fokker-Planck equation
\begin{equation}
-{\partial \over \partial t} \rho({\bf x},t)
=-{1\over 2}\nabla^2\rho({\bf x},t)
+\nabla_i\Bigl [G_i({\bf x})\rho({\bf x},t)\Bigr ]=L\rho({\bf x},t),
\label{FK}
\end{equation} 
correctly to second order. The reason for this is clear. In the limit
when the trial function is the exact ground state wave function 
$\phi({\bf x})\rightarrow \psi_0({\bf x})$, 
the local energy is the exact ground state energy, which is just
a constant. The convergence of the DMC algorithm would then coincide with the
convergence of the Langevin algorithm for simulating the Fokker-Planck equation.
Thus in order to have a second order DMC algorithm\cite{chin90}, 
one must have a second order 
Langevin algorithm, for example, by approximating 
\begin{equation}
{\rm e}^{-\epsilon L}={\rm e}^{-\epsilon(T+D)}\approx 
{\rm e}^{-{1\over 2}\epsilon T }
{\rm e}^{-\epsilon D}
{\rm e}^{-{1\over 2}\epsilon T }.
\end{equation}
 
This idea of operator factorization seemed promising for 
generating higher order DMC algorithms. However, 
Suzuki\cite{suzuki91} proved in 1991 that, beyond second order, 
it is impossible to factorize 
\begin{equation}
\exp[\epsilon (A+B)]
=\prod_{i=1}^N\exp[a_i\epsilon A]\exp[b_i\epsilon B] 
\label{facab}
\end{equation}
without having some coefficients $a_i$ 
and $b_i$ being negative. Since 
${\rm e}^{-a_i\epsilon T}$ is the diffusion kernel, a negative $a_i$ 
would imply a diffusion process backward in time, which is impossible
to simulate. Thus higher than second order DMC algorithms cannot be based on
obvious factorizations of the form (\ref{facab}).

In this work, we show how to derive a number of distinct quartic DMC algorithms 
by factorizing the operator ${\rm e}^{-\epsilon(T+D+E_L )}$
to fourth order with positive coefficients. We first review how each factorized 
operator can be simulated in Section II, followed by a derivation of a 
fourth order DMC algorithm in Section III. The backbone of this algorithm is
a 4th order Langevin algorithm, which is of importance in its own right.
In Section IV we examine the working
details of this algorithm and check its quartic convergence on various
systems including the practical case of liquid Helium. In Section V we discuss
alternative quartic algorithms by considering the unrestricted factorization of  
${\rm e}^{-\epsilon(T+D+E_L )}$. The convergences of two alternative 4th 
algorithms are also tested on liquid helium.
Our conclusions and suggestions for
future work are contained in Section VI.    

\section {Simulating the Basic Operators}

The method of operator factorization depends on the fact that each component
factor can be simulated exactly or to the required order. 
The effect of ${\rm e}^{-\epsilon T}$ on $\rho({\bf x},t)$ is to evolve
the latter forward in time according to the {\it diffusion} equation
\begin{equation}
-{\partial\over{\partial t}} \rho({\bf x},t)
=-{1\over 2}\nabla^2\rho({\bf x},t).
\label{tterm}
\end{equation}
For a set of points $\{x_i\}$ distributed according to $\rho({\bf x},t)$,
this can be exactly simulated by 
updating each point according to 
\begin{equation}
x_i^\prime=x_i+\sqrt{\epsilon}\,\xi_i,
\label{stterm}
\end{equation}
where $\{\xi_i\}$ is a set of Gaussian distributed random numbers with zero 
mean and unit variance. The operator ${\rm e}^{-\epsilon D}$ evolves $\rho({\bf x},t)$
forward in time according to the {\it continuity} equation
\begin{equation}
-{\partial\over{\partial t}} \rho({\bf x},t)
=\partial_i [G_i({\bf x})\rho({\bf x},t)],
\label{tdterm}
\end{equation}
where ${G_i}({\bf x})\rho({\bf x},t)={J_i}({\bf x})$ is the particle current 
density with drift velocity field ${G_i}({\bf x})$. This
can also be exactly simulated by setting 
\begin{equation}
x_i^\prime=x_i(\epsilon),
\label{trterm}
\end{equation}
where $x_i(\epsilon)$ is the exact trajectory determined by
\begin{equation}
{{d{\bf x}}\over{dt}}={\bf G}({\bf x}),
\label{traject}
\end{equation}
with the initial condition $x_i(0)=x_i$. In practice, one can only
solve this trajectory equation to the required order of accuracy.
The operator ${\rm e}^{-\epsilon E_L}$ evolves 
$\rho({\bf x},t)$
forward in time according to the {\it rate} equation
\begin{equation}
-{\partial\over{\partial t}} \rho({\bf x},t)
=E_L({\bf x})\rho({\bf x},t).
\label{eterm}
\end{equation}
The exact solution
\begin{equation}
\rho({\bf x},t+\epsilon)
={\rm e}^{-\epsilon E_L({\bf x})}\rho({\bf x},t)
\label{rterm}
\end{equation}
can be simulated by updating the {\it weight} $W_k$ associated with the 
configuration ${\bf x}_k$ by
\begin{equation}
W_k^\prime
={\rm e}^{-\epsilon \left [E_L({\bf x}_k)-E\right ]}W_k.
\label{wterm}
\end{equation}
A uniform constant $E$ is usually added to keep the weights near unity. 

There are various methods\cite{kalos62,hetherington84,nightingale88,cerf95}
of keeping track of weights, the original and simplest 
method\cite{kalos62} 
is just to replicate the configuration ${\bf x}_i$ on the average
${\rm e}^{-\epsilon [E_L({\bf x}_i)-E]}$ times. We use a method which is 
intermediate between that of \cite{hetherington84}  
and \cite{nightingale88}. Our algorithm is, however, independent of 
any specific method of weight tracking. 
  
Thus the second order factorization of
\begin{equation}
{\rm e}^{-\epsilon(T+D+E_L )}\approx 
{\rm e}^{-{1\over 2}\epsilon E_L }
{\rm e}^{-{1\over 2}\epsilon T }
{\rm e}^{-\epsilon D }
{\rm e}^{-{1\over 2}\epsilon T }
{\rm e}^{-{1\over 2}\epsilon E_L }+o(\epsilon^3),
\end{equation}
leads to the following second order DMC algorithm:
\begin{eqnarray}
y_i=&&x_i+\xi_i\sqrt{\epsilon/2}\,,\nonumber\\
x^\prime_i=&& y_i(\epsilon)+\xi^\prime_i\sqrt{\epsilon/2},
\label{altas}
\end{eqnarray}
where the final position ${\bf x}^\prime$ is to be weighted by 
\begin{equation}
W^\prime={\rm e}^{-{1\over 2}\epsilon
\left [E_L({\bf x})+E_L({\bf x}^\prime)-2E\right ]},
\end{equation}
and where each $y_i(\epsilon)$ needs to be solved at least to second order by 
any convenient method. Following \cite{chin90} this algorithm will be
referred to as DMC2b.  

\section {A Fourth Order Algorithm}

For a fourth order factorization of $\exp[\epsilon (A+B)]$ 
with positive coefficients, Suzuki\cite{suzuki91} has
shown that it is necessary to retain as a factor,
the exponential of either double 
commutators $[A,[B,A]]$ or $[B,[A,B]]$.
Recently, Chin\cite{chin97} has derived three 
such factorization schemes, two of which were also found previously by 
Suzuki\cite{suzuki96}. To decompose 
${\rm e}^{-\epsilon(T+D+E_L )}={\rm e}^{-\epsilon(L+E_L )}$ to fourth order,
one possibility is to keep the Langevin operator $L$ intact. In this case, 
the double commutator 
\begin{equation}
\left[E_L,\left[ L , E_L \right]\right] 
= \left[E_L,\left[ T , E_L \right]\right]
= (\partial_iE_L)(\partial_i E_L),
\end{equation}
is the square of the gradient of the local energy, which is a 
manageable coordinate function. Since the Langevin operator 
is complicated to simulate, we must choose a fourth order 
factorization of
${\rm e}^{-\epsilon(L+E_L )}$ 
which minimizes the appearance of $L$. 
We choose the following factorization as given by Refs.\cite{chin97,suzuki96}, 
\begin{equation}
{\rm e}^{-\epsilon \left( L + E_L\right)} =
{\rm e}^{-{1\over 6}\epsilon  E_L}
{\rm e}^{-{1\over 2}\epsilon L}
{\rm e}^{-{2\over 3} \epsilon \tilde{E_L}}
{\rm e}^{-{1\over 2}\epsilon L}
{\rm e}^{-{1\over 6}\epsilon E_L}
+o(\epsilon^5),
\label{dmc41}
\end{equation}
with $\tilde{E_L}$ given by
\begin{equation}
\tilde{E}_L = E_L + {1\over 48} \epsilon^2 \left[ E_L,\left[L,E_L\right]\right]
= E_L + {1\over 48} \epsilon^2 \left\vert \nabla E_L \right\vert^2.
\label{dmc4p1b}
\end{equation}
Thus to the extend that the local energy $E_L({\bf x})$ is a smooth function,
the double commutator correction will be negligible. 

The weights
in (\ref{dmc41}) have a simple structure. If ${\bf x}_0$ 
is the initial configuration, ${\bf x}_{1/2}$ the Langevin evolved 
configuration time step $\epsilon/2$ later, and ${\bf x}_1$ the Langevin
evolved configuration a time step $\epsilon/2$ later still, then we assign
the final configuration ${\bf x}_1$ a weight of
\begin{equation}
W_1={\rm e}^{-\epsilon
\left [{1\over 6}E_L({\bf x}_1)+{2\over 3}\tilde{E}_L({\bf x}_{1/2})
+{1\over 6}E_L({\bf x}_0)
-E\right ]}.
\end{equation}
 
The demanding part of this DMC algorithm is the simulation the Fokker-Planck
equation (\ref{FK}). The resulting Langevin algorithm is an important 
simulation algorithm with numerous applications in statistical and chemical 
physics\cite{risken}. Since we have recently given a detailed derivation
of a fourth order Langevin algorithm\cite{forbertchin00}, 
we will be brief here in summarizing its essential features.  
To obtain a fourth order Langevin algorithm, we again seek to decompose
${\rm e}^{-\epsilon L}={\rm e}^{-\epsilon(T+D)}$ to fourth order. In this case,
we keep the double commutator $[D,[T,D]]$, which is at most a second order 
differential operator, and factorize the Fokker-Planck operator as\cite{chin97}, 
\begin{equation}
{\rm e}^{-\epsilon L}={\rm e}^{-\epsilon \left( T + D\right)} =
{\rm e}^{-{1\over 2}(1-{1\over\sqrt{3}})\epsilon T}
{\rm e}^{-{1\over 2}\epsilon D }
{\rm e}^{-{1\over\sqrt{3}}\epsilon\tilde{T} }
{\rm e}^{-{1\over 2}\epsilon D }
{\rm e}^{-{1\over 2}(1-{1\over\sqrt{3}})\epsilon T}
+o(\epsilon^5),
\label{dmc4p2}
\end{equation}
\begin{equation}
{1\over\sqrt{3}}\tilde{T}={1\over\sqrt{3}}T + {\epsilon^2\over 24}
(2-\sqrt{3})
[D,[T,D]]={1\over\sqrt{3}}T+ {\epsilon^2\over 24}(2-\sqrt{3})
[\partial_i \partial_j f_{i,j} + \partial_i v_{i}],
\label{ttid}
\end{equation}
where subscripts indicate partial differentiations,
and	
\begin{eqnarray}
f_{i,j}\,&&\equiv 
2 S_{i,k} S_{j,k} - S_{i,j,k} S_{k} \\
v_i\,&&\equiv - {1\over 2} \left(
2 S_{i,j,k} S_{j,k} + S_{i,j} S_{j,k,k} - S_{i,j,k,k} S_{j} \right).
\label{fdef}
\end{eqnarray}
By appropriate normal ordering, the double commutator term can
be regarded as a non-uniform Gaussian random walk. However, in order
to be able to sample the non-uniform Gaussian in cases where 
$f_{i,j}$ has negative eigenvalues, we implement the normal ordering
as follows so that the full covariance matrix is always positive
definite in the limit of small $\epsilon$, 
\begin{eqnarray}
\exp&&\left( {\epsilon\over\sqrt{3}}\tilde{T} \right)
=
\exp\left( {\epsilon\over 2\sqrt{3}} T\right)
{\cal N} \left\{
\exp\left[
 {\epsilon^3\over 24}\left(2 - \sqrt{3}\right)
\left(\partial_i \partial_j f_{i,j} +
\partial_i v_i \right) \right]
\right\}
\exp\left( {\epsilon\over 2\sqrt{3}} T\right)
\nonumber\\
&&=																	   
{\cal N} \left\{
\exp\left[ {\epsilon\over 2\sqrt{3}} 
\left(- {1\over 2} \partial_i \partial_j \delta_{i,j}\right)
 +{\epsilon^3\over 24}\left(2 - \sqrt{3}\right)
\left(\partial_i \partial_j f_{i,j} +
\partial_i v_i \right) \right]
\right\}
\exp\left( {\epsilon\over 2\sqrt{3}} T\right),
\end{eqnarray}
where ${\cal N}$ denotes the normal ordering of all derivative operators
to the left.
Factorization (\ref{dmc4p2}) can now be simulated as
\begin{eqnarray}
w_i &=& x_i +
     \xi_i\sqrt{{\epsilon\over 2} \left(1-{1\over\sqrt{3}}\right)},
     \nonumber \\
y_i &=& w_i(\epsilon/2)
     +\xi^\prime_i \sqrt{\epsilon\over 2 \sqrt{3}},\nonumber \\
z_i &=&
     y_i    - {\epsilon^3\over 24} \left( 2 - \sqrt{3} \right) v_i({\bf y})
	 + \sqrt{\epsilon\over 2 \sqrt{3}} 
   \left[
   \delta_{i,j} + {1\over 2}
   \left({1\over\sqrt{3}}-{1\over 2} \right)\epsilon^2 f_{i,j}({\bf y})
   \right]
   \xi_j^{\prime\prime},\nonumber\\
x_i^\prime &=& z_i(\epsilon/2) 
      +\xi^{\prime\prime\prime}_i
	  \sqrt{{\epsilon\over 2} \left(1-{1\over\sqrt{3}}\right)},
\label{dmc4al} 
\end{eqnarray}
where $\xi_i$ to $\xi_i^{\prime\prime\prime}$ are four sets of
independent Gaussian random numbers with zero mean and unit variance.
Here, the two trajectory equations $w_i(\epsilon/2)$, $z_i(\epsilon/2)$
must be solved correctly to at least fourth order. Empirically one observes 
that the more accurately one solves the trajectory equation, the smaller is the
fourth order error coefficient. However, in practice one must weight improved
convergence, which allows larger time steps to be used, against greater 
computational effort. In the present case, we solve the trajectory 
equation by the standard 4th order
Runge-Kutta algorithm.

Eq.(\ref{dmc41}) is our basic 4th order DMC algorithm and will be referred to
as DMC4. We will first explore
its workings in some detail before considering alternative algorithms. 

\section {Applying the Fourth Order Algorithm}

We begin by verifying that DMC4 is indeed quartic by solving the 3-D 
harmonic oscillator
\begin{equation}
H=-{1\over 2}\nabla^2+{1\over 2}r^2
\label{harh}
\end{equation}
both analytically with the help of Mathematica and by direct Monte 
Carlo simulation. The trial function used is
\begin{equation}
\phi({\bf r})=\exp(-{1\over 2}\alpha r^2)
\label{harw}
\end{equation}
with a deliberate poor choice of the trial parameter $\alpha=1.8$.
In Fig.1 we plot the ground state energy from the mixed expectation 
\begin{equation}
E={{\langle\phi|H|\psi_0\rangle}\over
{\langle\phi|\psi_0\rangle}	}
\label{emix}
\end{equation}
as a function of the step size $\epsilon$ used. The lines are analytical
functions from Mathematica and the plotting symbols are Monte Carlo 
simulation results. We have
included one first order, two second order, and one first order rejection
DMC algorithm for comparison. The detail descriptions of DMC1, DMC2a and 
DMC2b can be found in Ref.\cite{chin90}. The rejection algorithm
uses a first order Langevin algorithm together with a generalized Metropolis 
acceptance-rejection step so that the square of the trial function is exactly
sampled at all step sizes\cite{reynolds82}. For this case, we have no analytical 
result and the plotted line is just a 6th order polynomial fit. The 4th
order algorithm is indeed quartic, as can be verified
analytically. It is distinct from lower order algorithms even in Monte Carlo
simulations.

We next test DMC4 by solving the 3-D Morse potential with Hamiltonian,
\begin{equation}
H=-{1\over 2}\nabla^2			 
+D_e\Bigl [{\rm e}^{-2\alpha(r-r_0)}-2{\rm e}^{-\alpha(r-r_0)}\Bigr ],
\label{morh}
\end{equation}
with $D_e=50$, $r_0=1$ and $\alpha=10$. These values ensure that 
the ground state is high up in the potential and that the ground
state wave function is not well approximated by a Gaussian. We use a trial
function of the form
\begin{equation}
\phi({\bf r})=\exp(-ar-br^{-3}),
\label{morw}
\end{equation}
with $a=15.29$, $b=6.82$ and variational energy -11.1774. This is to be
compared with the exact ground state energy of -12.5$\,$. The convergence 
of various DMC algorithms are compared in Fig.2. The quartic convergence 
of DMC4 is again verified. Its convergence is clearly distinct 
from lower order results and is nearly flat. In this case, we have no 
analytical results and all lines are just least square fits to the data. 

To demonstrate that DMC4 can be used to solve realistic 
physical problems, we use it to solve for ground state properties of 
bulk liquid helium described by the many-body Hamiltonian
\begin{equation}
H=\sum_i-{\hbar^2\over 2m}\nabla^2_i
+\sum_{i<j}V(r_{ij}),
\label{helh}
\end{equation}
where $\hbar^2/m=12.12 \AA^2 K$ with potential $V$ determined by
Aziz {\it et al} \cite{aziz79}. Instead of the usual McMillan trial function,
we use a trial function of the form,
\begin{equation}
\phi({\bf x})=\prod_{i<j}\exp\{-{\rm ln}(2)\exp[-(r_{ij}-c_0)/d_0]\}.
\label{chinw}
\end{equation}
With $c_0=2.8\AA$ and $d_0=0.48\AA$, this trial function gives a slightly better
energy of 5.886(5) K/particle.
Since the standard calculation details\cite{gfmc} are well known, we will
just describe the results as summarized in Fig.3. Again, the convergence of our
4th order algorithm is clearly quartic. The extrapolated values are -7.114(2) $K$
for our 4th order algorithm and -7.111(2) $K$ for the second order algorithm DMC2a.
Both are in agreement with Boronat and Casulleras's\cite{boronat94}
second order DMC result of -7.121(10) $K$.
Note that very large steps can be used with algorithm DMC4, roughly ten
times as large as those of second order algorithms.

In Fig.4, we show the resulting radial density distribution $g(r)$ from our
4th order calculation at $\epsilon=0.1$ and $\epsilon=0.2$. The distribution is
virtually unchanged even at these large time steps and both are in excellent 
agreement with the experimental distribution of 
Svensson {\it et al.} \cite{svensson}.

In Fig.5 we show DMC4's thermalization toward the exact ground 
state from the variational trial wave function. Starting from the initial 
variational energy, we plot the population averaged energy as a function of 
iterated time for various time step sizes. This plot shows that each iteration
of the algorithm at $\epsilon=0.2$ is indistinguishable from multiple iterations
at smaller time steps having the same time interval. Moreover, it demonstrates 
that the algorithm converges to the ground state inversely
proportional to the step size used, up to $\epsilon=0.2$. That is,
only five iterations are needed at $\epsilon=0.2$ to reach the exact ground 
state near $t=1.0$ and 20 iterations at $\epsilon=0.05$, etc.. Thus our 4th order 
algorithm can project out the ground state with ten times fewer updates than a 
second order algorithm.

More important than the thermalization time is the observable correlation time.
In a Monte Carlo calculation, it is highly desirable to have uncorrelated 
configurations for an accurate estimate of the statistical errors. 
The correlation coefficient for an observable $O$ is defined by
\begin{equation}
c_O(\Delta t)={{<O(t+\Delta t)O(t)>-<O(t)>^2}
              \over{<O(t)O(t)>-<O(t)>^2}}.
\label{core}
\end{equation}
In Fig.6, we show the ground state energy correlation function of
liquid helium as computed by our 4th order algorithm. The correlation time is
roughly $\Delta t\approx 1.5$, at which point the correlation coefficient
dropped to zero. This plot shows that the correlation time depends only on
the total time separation. Thus if the algorithm remains accurate
at large time steps, then fewer iterations are needed to produce 
uncorrelated configurations. 

In implementing the fourth order Langevin algorithm, we used the standard 
fourth order Runge-Kutta algorithm to solve the trajectory equation (\ref{traject}).
When the step size is large, the 4th order error in the Runge-Kutta algorithm
can greatly overshadow the intrinsic 4th order step size error of the 
Langevin and that of the DMC algorithm, causing both to fail prematurely. 
To guard against this from happening, 
we monitor the difference between the results of the fourth order 
Runge-Kutta and its embedded second order algorithm.
If the square of this difference is larger than some tolerance, say $0.01$ ,
we recalculate the trajectory twice at half the time step size.
Even at the largest step size used, only a few percent of
trajectories need to be recalculated, incurring only a small additional
overhead. This additional effort greatly extended the flatness of the
convergence curve as shown in Fig.3. The total calculation time running on a
single processor of an Origin 2000 is 51 hours, about 5 times as long the second
order algorithm. 

\section {Alternative Fourth Order Algorithms}

Our DMC4 algorithm (\ref{dmc41}), which preserves the Fokker-Planck
operator $L$ intact, may not be the most efficient fourth order algorithm 
possible. Consider the limit in which the trial function approaches the exact
ground state, $\phi({\bf x})\rightarrow\psi_0({\bf x})$. In this ideal case 
the local energy becomes an irrelevant constant, $E_L({\bf x})\rightarrow E_0$,
and the algorithm is just  
\begin{equation}
{\rm e}^{-\epsilon \left( L + E_L\right)} \propto
{\rm e}^{-{1\over 2}\epsilon L}
{\rm e}^{-{1\over 2}\epsilon L},
\label{dmclim}
\end{equation}
which is the running of the 4th order Langevin algorithm 
twice, at half the time step. It seems plausible that one should be
able to derive a 4th order DMC algorithm which reduces to   
a single run of the 4th order Langevin algorithm in the same limit.

We are thus led to consider the general factorization, to fourth order, of a
three-operator exponential ${\rm e}^{-\epsilon \left(T+D+E_L\right)}$.
There are now 9 double commutators to be considered:
6 are the generalizations of the two operator case,
\begin{equation}
[T,[D, T]],\quad [D,[T,D]],\quad 
[D,[E_L, D]],\quad [E_L,[D,E_L]],\quad 
[E_L,[T,E_L]],\quad [T,[E_L,T]],\nonumber
\end{equation}
and three new ones related by the Jacobi identity,
\begin{equation}
[T,[D, E_L]]+[D,[E_L,T]]+[E_L,[T, D]]=0.
\end{equation}
Thus only two of the last three commutators are independent. Note also that 
for the present form of the operators, $[E_L,[D, E_L]]=0$.
We have examined all these commutators in the case of liquid helium to
determine which one is doable and can be kept.    
To explore the many possible factorizations, we have devised a 
Mathematica program to combine the exponential of operators 
symbolically. With the help of this program, we have explored an extensive
list of distinct 4th order algorithms. Since there are many operators
in each such factorization, it is too cumbersome to 
write out the explicit exponential form. Moreover, since the factorization 
will always be left-right symmetric, there
is no need to repeat operators on the left side. In the following, we will
only indicate the exponential operators symbolically beginning with the 
{\it central} one and list only operators to the {\it right}. 
For example, algorithm DMC4 (\ref{dmc41}) will be 
denoted as
\begin{eqnarray}
T+D+E_L && \approx
{{2\over 3}\tilde{E_L}}
{+{1\over 2} L}
{+{1\over 6} E_L} \nonumber \\
        && \approx 
{{2\over 3}\tilde{E_L}}
+{1\over 2}\left[{{1\over 2}(1-{1\over\sqrt{3}}) T}
+{{1\over 2} D }
+{{1\over\sqrt{3}}\tilde{T} }
+{{1\over 2} D }
+{{1\over 2}(1-{1\over\sqrt{3}}) T}\right]
{+{1\over 6} E_L}
.
\label{newnot}
\end{eqnarray}
Each update of this algorithm requires the evaluation of, in 
decreasing order of computational complexity,
4$D$'s, 2$\tilde T$'s, 1$\tilde E_L$, 1$E_L$ and 4$T$'s.
(The last $E_L$ from the last update can be used as first $E_L$ of the
current update.)   
Since $D$ is the most computationally intensive 
operator, followed by $\tilde T$, $\tilde E_L$, etc., we would like to 
minimize their appearance in that order. Below, we will describe two
alternative algorithms that are computationally more economical than
DMC4 in solving for the ground state of liquid helium.
 
One possible 4th order algorithm is to retain the 
same double commutators $[D,[T,D]]$ and $[E_L,[T,E_L]]$ as in DMC4, but allow
$L=T+D$ to be broken up: 
\begin{equation}
T+D + E_L \approx
{1\over 3}\tilde T
+{  1\over 6}D 
+{3\over 8}E_L 
+{  1\over 6}T 
+{  1\over 3}D 
+{  1\over 6}T 
+{1\over 8}\tilde{E_L}.
\label{dmc42}
\end{equation}
Here, $\tilde T$ and $\tilde E$ given by
\begin{equation}
\tilde{T}=T + {\epsilon^2\over 72}
[D,[T,D]],
\label{ttil}
\end{equation}
\begin{equation}
\tilde{E_L}=E_L + {\epsilon^2\over 12}
[E_L,[T,E_L]].
\label{etill}
\end{equation}
This algorithm requires 
4$D$'s, but only 1$\tilde T$, 1$\tilde E_L$, 2$E_L$'s and 4$T$'s. We will denote
this algorithm as DMC4a. This algorithm is roughly 10\% faster DMC4 and its
quartic convergence is clearly demonstrated in Fig.\ref{fthree}. However, its 
convergence range is only about half of DMC4. The actual running time for 
this algorithm is 46 hours. 
 
To reduce the number of $D$ operators, one must pay the price of
retaining additional double commutators. We will refer to the following
algorithm with only two $D$ operators as DMC4b:
\begin{equation}
T+D + E_L \approx
{1\over {\sqrt{3}}}a_0\tilde T
+{1\over {2\sqrt{3}}}{E_L} 
+{1\over {2\sqrt{3}}}(1-a_0)T 
+{  1\over 2}D 
+{  1\over 4}c_0T 
+{  1\over 2}c_0E_L
+{  1\over 4}c_0T, 
\label{dmc4b}
\end{equation}
where $a_0=1/\sqrt{1+\sqrt{3}}$, $c_0=1-1/\sqrt{3}$ and
\begin{eqnarray}
{1\over\sqrt{3}} a_0\tilde{T}={1\over 2\sqrt{3}}a_0T 
&& + {\epsilon^2\over 24}
\left[
(2-\sqrt{3})\Bigl([D,[T,D]]+[D,[E_L,D]]\Bigr)
+(c_0-{a_0\over {\sqrt{3}}})[E_L,[T,E_L]]
\right] \nonumber \\
&&+{1\over 2 \sqrt{3}}a_0T.
\label{tttid}
\end{eqnarray}
The additional commutator
\begin{equation}
[D,[E_L,D]]=-G_i\nabla_i[G_j\nabla_j E_L({\bf x})],
\label{ded}
\end{equation}
is a calculable function involving higher derivative of $G_i$ and $E_L$.
In this algorithm, we have placed all the double commutators at 
the center so that they are
evaluated only once per update. This is done by splitting 
${1\over\sqrt{3}}a_0T\rightarrow
{1\over 2\sqrt{3}}a_0T+...{1\over 2\sqrt{3}}a_0T$ in Eq.(\ref{tttid}), 
meaning that we first 
do half of required Gaussian random walk, evaluate all double commutators, 
then complete the remaining half of the Gaussian walk including the
effect of $[D,[T,D]$ as it is done in the Langevin algorithm. 
The ubiquitous irrational coefficients are roots of quadratic equations
which force unwanted double commutators to vanish. One can check by
inspection that as $\phi({\bf x})\rightarrow\psi_0({\bf x})$
and $E_L({\bf x})\rightarrow E_0$, (\ref{dmc4b}) reduces to just
the 4th order Langevin algorithm (\ref{dmc4p2}). 

The results of this algorithm for liquid helium are shown in 
Fig.\ref{fthree} as asterisks. The ground state energy is correctly obtained, 
but because these higher derivatives are rapidly varying functions, we have not
been able to stabilize the population of weights beyond 
$\epsilon\approx 0.05$. While this algorithm may not work as well as DMC4 and
DMC4a for liquid helium, its economy of requiring only two trajectories per 
update may be of utility in other applications. The calculation time for data
points shown is 36 hours.    

\section{Conclusions}

In this work, we have derived a number of distinct fourth order DMC algorithms
by factorizing the imaginary time Schr\"odinger evolution operator to 
fourth order. This is a notable advance in algorithm development, 
made possible only by the recent progress in understanding positive 
coefficient operator factorization. Our work illustrates a global view of 
algorithms as products of factorized operators. Such a perspective gives order
and insight into the working of the Diffusion Monte Carlo algorithm.
It would have been very difficult to derive such a high order algorithm 
without such a conceptual structure. We have further demonstrated 
the practicality of these algorithms by using 
them to solve for the ground state of liquid helium. 
The quartic convergences of 
DMC4 and DMC4a have been verified and both yielded
ground state energy and radial density distribution
in excellent agreements with experiment.
Despite the fact that these algorithms are rather complicated to program
requiring higher order derivatives, they allow very large step sizes to
be used, virtually eliminate the time step size error and greatly reduce 
statistical correlations between successive updated configurations. 

Since this
is only the first demonstration of quartic algorithms, there
is room for further improvements. For example, we have shown how
the factorization of a three-operator exponential can lead to a number of
distinct four order DMC algorithms. A more systematic categorization of 
various fourth order factorizations would help in obtaining the most efficient
algorithm. Secondly, the retainment of some double commutators is necessary, 
however, it has not been studied in detail where they should be placed so as
to minimize the 4th order error coefficient or computational effort.
It is observed that the step size convergence curve is flatter when the  
trajectory equation is solved more exactly. In this work, we have only used
the 4th order Runge-Kutta algorithm in solving for the deterministic 
trajectory. Future study may explore the effects of using
alternative numerical methods, such as
symplectic algorithms\cite{forest}, for solving the trajectory equation.

\acknowledgements
This research was funded, in part, by the U. S. National Science Foundation 
grants PHY-9512428, PHY-9870054 and DMR-9509743.


\ifpreprintsty\newpage\fi
\ifpreprintsty\newpage\fi
\begin{figure}
\noindent
\vglue 0.2truein
\hbox{
\vbox{\hsize=7truein
\epsfxsize=6truein
\leftline{\epsffile{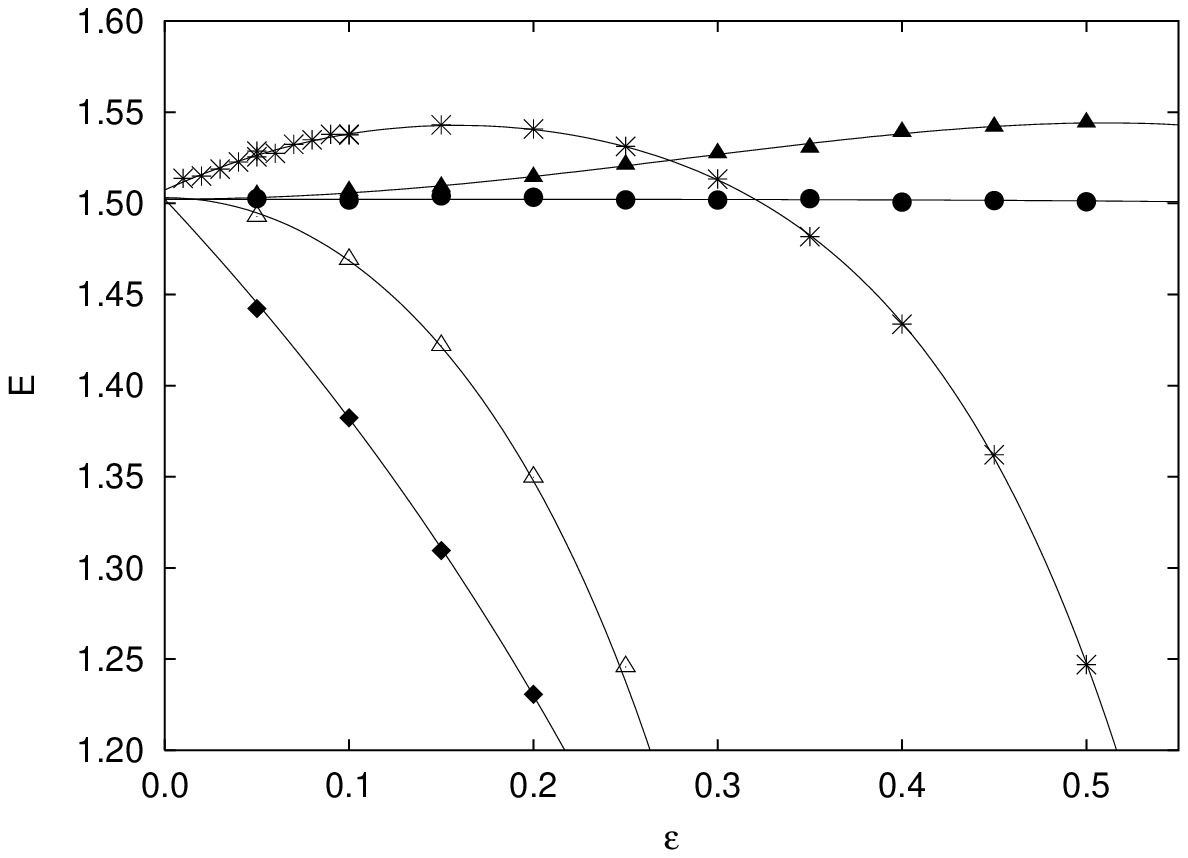}}
}
}
\vglue 0.3truein
\caption{The convergence of the ground state energy of a 3-D harmonic
oscillator as a function of the time step size $\epsilon$. The diamonds
are first order DMC1 simulation results. The filled and open triangles are 
second order DMC2a and DMC2b results. The asterisks indicate results of a 
linear algorithm with rejection. The filled circles are DMC4 results,
Eq.(\ref{dmc41}).
The errorbars are smaller than the symbol size. 
The various lines are the corresponding exact analytical results except in the
case of the rejection algorithm. For the latter case the line is just a
6th order polynomial fit to the simulation data.
 }
\label{fone}
\end{figure}
\ifpreprintsty\newpage\fi
\begin{figure}
\noindent
\vglue 0.2truein
\hbox{
\vbox{\hsize=7truein
\epsfxsize=6truein
\leftline{\epsffile{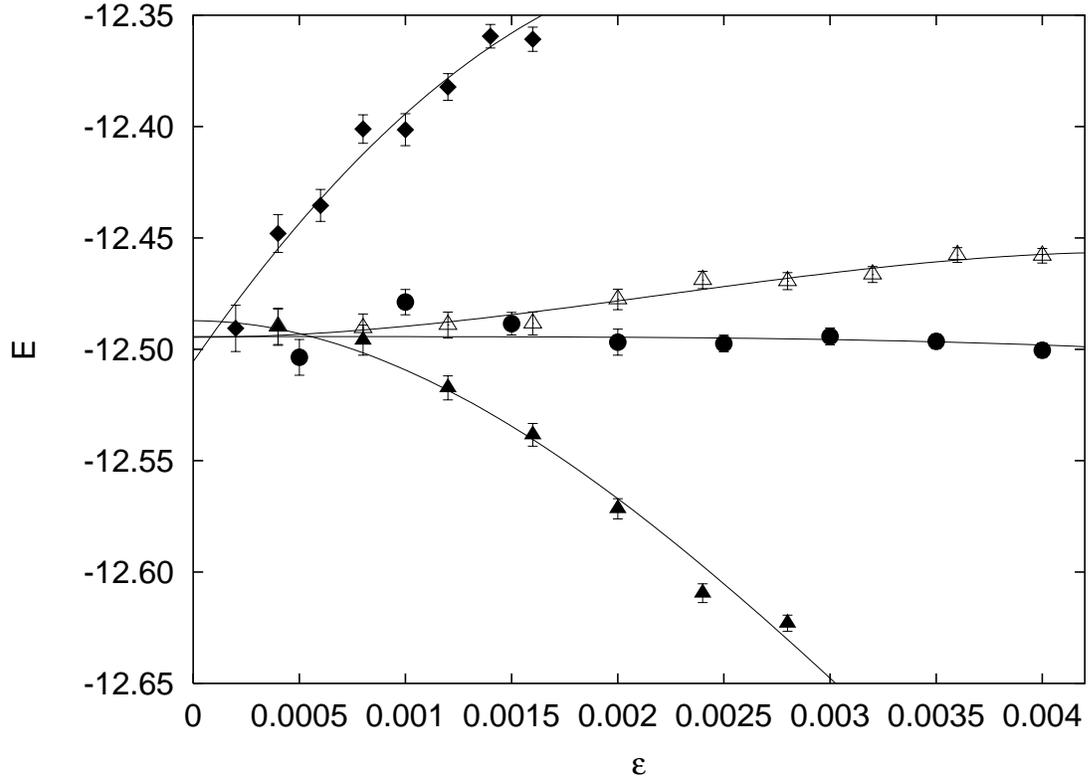}}
}
}
\vglue 0.3truein
\caption{The convergence of the ground state energy of a 3-D Morse
oscillator as a function of the time step size $\epsilon$. The diamonds
are first order DMC1 simulation results. The filled and open triangles are 
second order DMC2a and DMC2b results. The filled circles are the 4th order
results of DMC4. 
The various lines are least square fits to the simulation data.
The first order results are fitted with a parabola, the second order
results by a cubic polynomial, and the fourth order results by just
a constant plus a fourth order term in $\epsilon$.
}
\label{ftwo}
\end{figure}
\ifpreprintsty\newpage\fi
\begin{figure}
\noindent
\vglue 0.2truein
\hbox{
\vbox{\hsize=7truein
\epsfxsize=6truein
\leftline{\epsffile{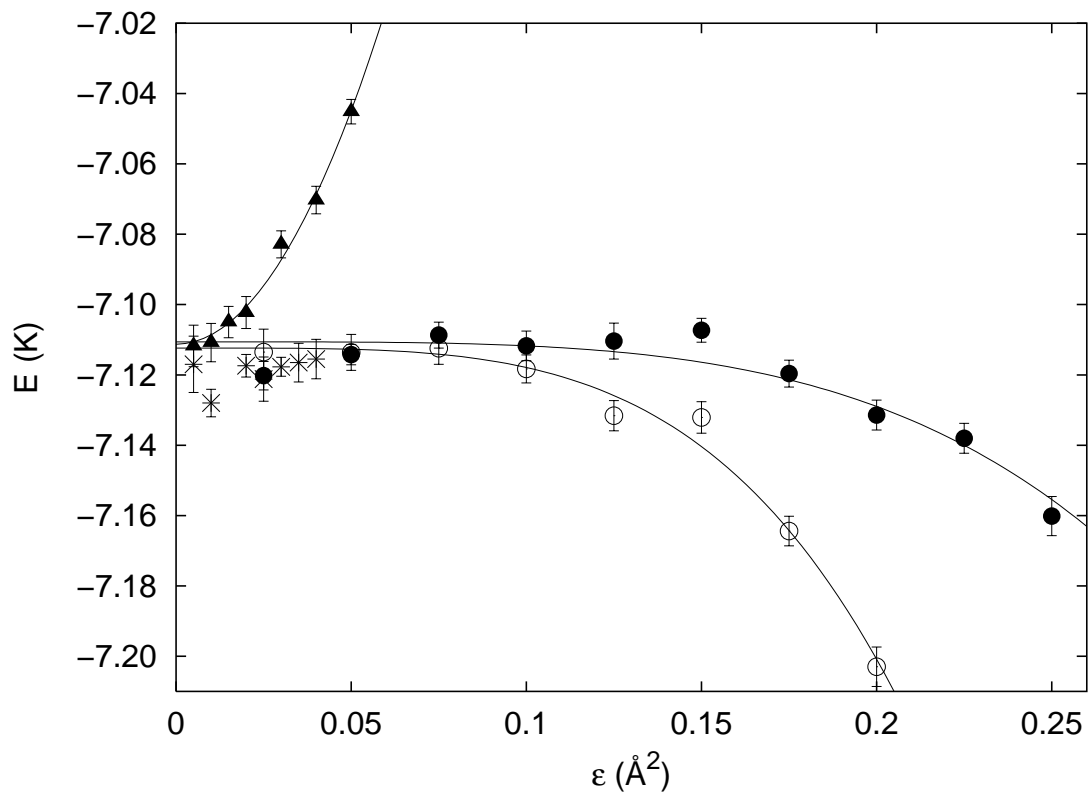}}
}
}
\vglue 0.3truein
\caption{The time step convergence of the ground state energy per particle 
for bulk liquid helium in a 128 particle simulation. The solid circles are 
the result of our 4th order algorithm DMC4. The open circles and asterisk are 
for algorithm DMC4a and DMC4b respectively. For comparison, we also show 
as triangles, second order results from algorithm DMC2a. 
The lines are least square fits to the data.
}
\label{fthree}
\end{figure}
\ifpreprintsty\newpage\fi
\begin{figure}
\noindent
\vglue 0.2truein
\hbox{
\vbox{\hsize=7truein
\epsfxsize=6truein
\leftline{\epsffile{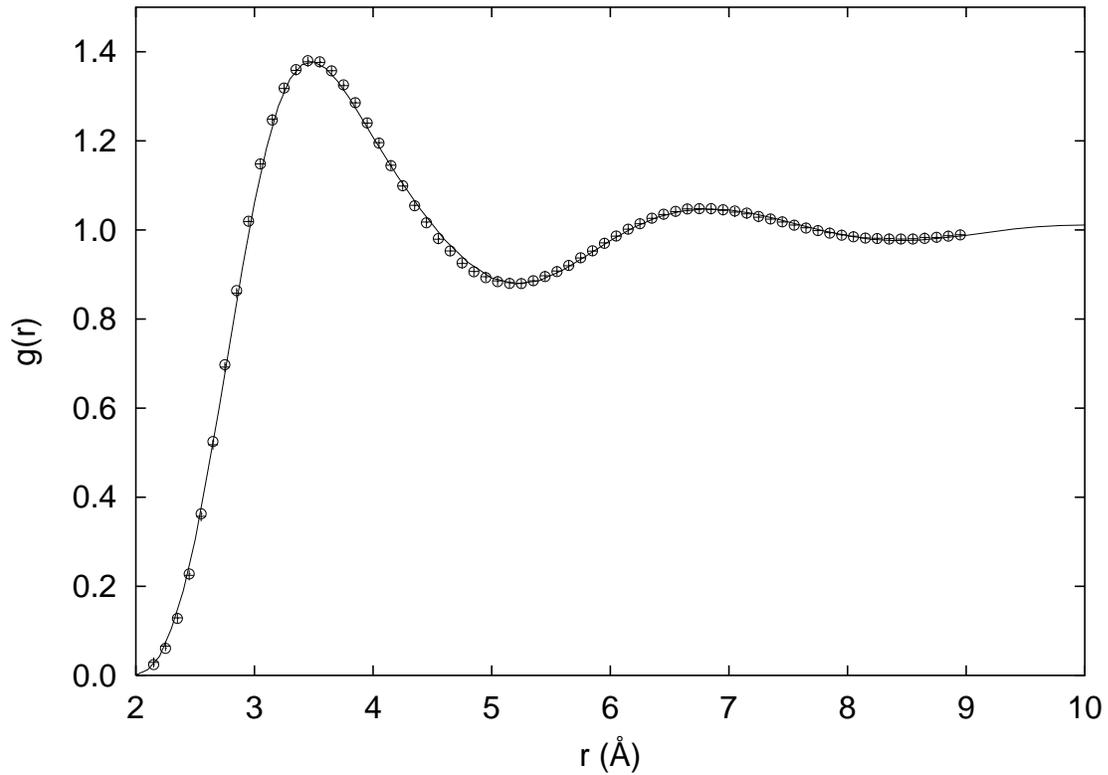}}
}							
}
\vglue 0.3truein
\caption{The radial density distribution of bulk liquid 
helium. The circles are DMC4 results at $\epsilon=0.1$
and the crosses are DMC4 results at $\epsilon=0.2$. The solid line
is the experimentally extracted $g(r)$ of 
Svensson {\it et al.}\protect\cite{svensson} at 1 $K$. 
}
\label{ffour}
\end{figure}
\ifpreprintsty\newpage\fi
\begin{figure}
\noindent
\vglue 0.2truein
\hbox{
\vbox{\hsize=7truein
\epsfxsize=6truein
\leftline{\epsffile{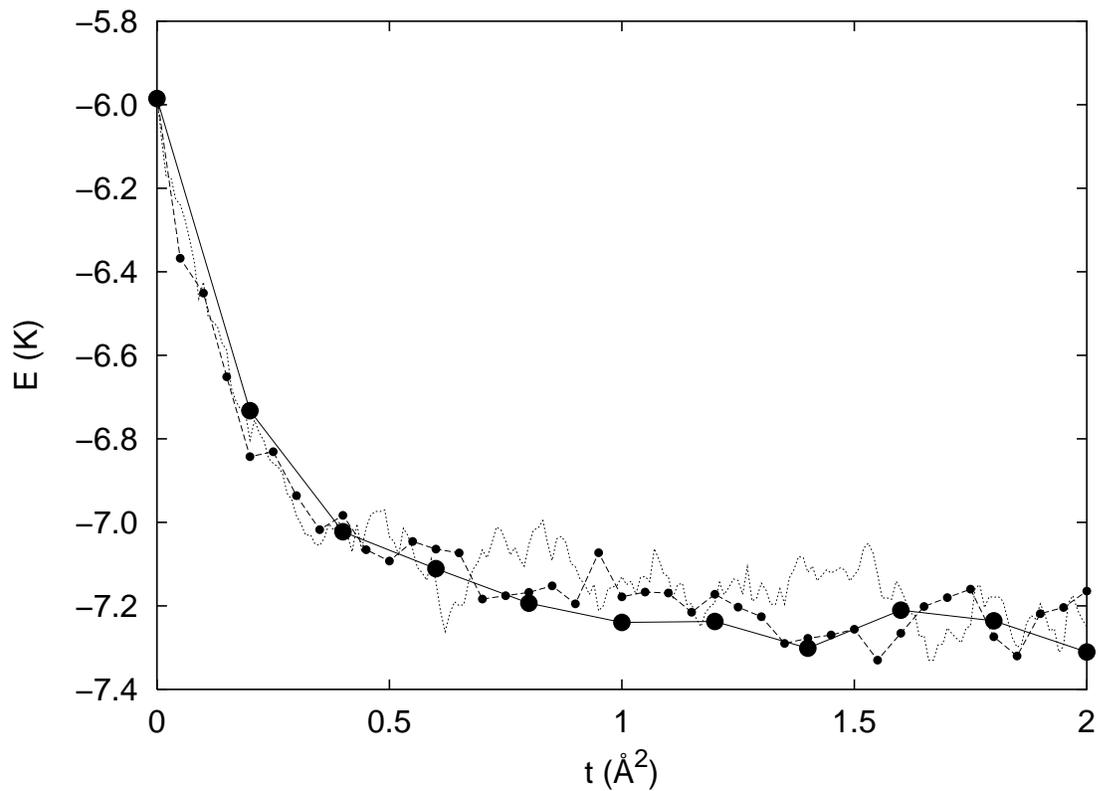}}
}
}
\vglue 0.3truein
\caption{The relaxation of liquid helium's ground state 
energy toward its exact value as simulated by DMC4 at 
three time step sizes. The large
circles are at $\epsilon=0.2$ and the small circles are at 
$\epsilon=0.05$. They are connected by straight line
segments to guide the eye. The dotted line corresponds to
results at $\epsilon=0.01$. 
}
\label{ffive}
\end{figure}
\ifpreprintsty\newpage\fi
\begin{figure}
\noindent
\vglue 0.2truein
\hbox{
\vbox{\hsize=7truein
\epsfxsize=6truein
\leftline{\epsffile{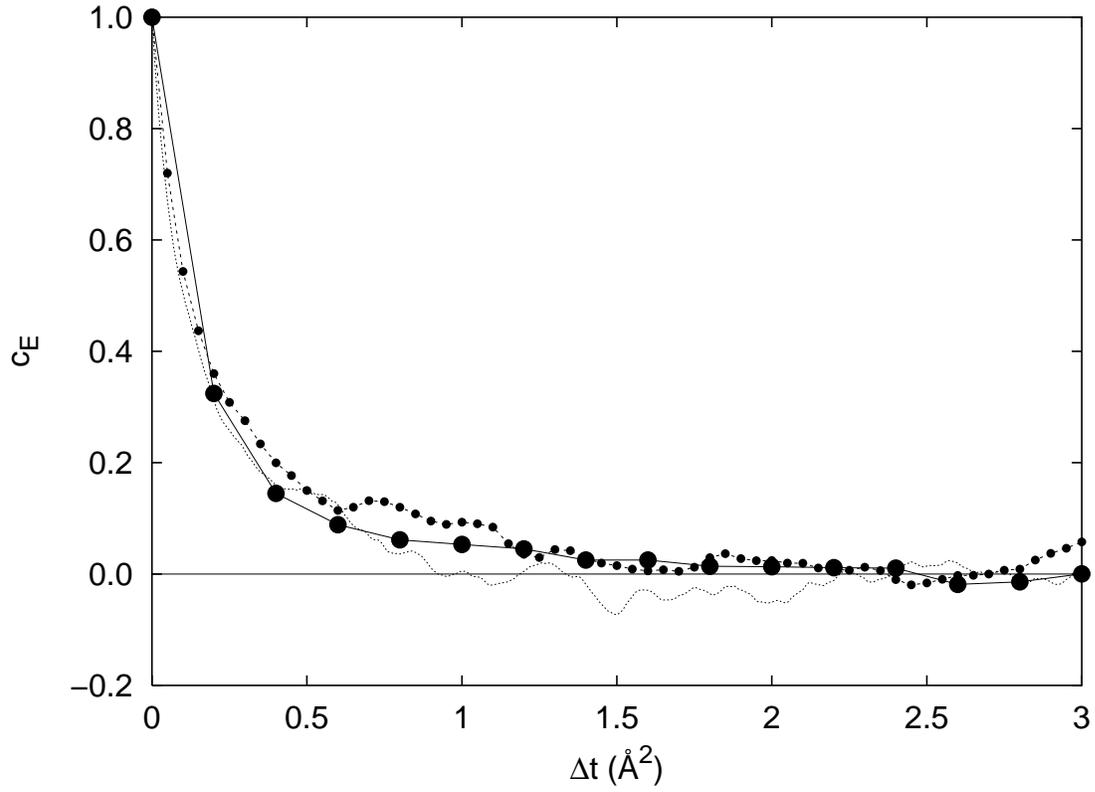}}
}
}
\vglue 0.3truein
\caption{The correlation coefficient function, Eq.(\ref{core}),
for the ground state energy of liquid helium as computed by DMC4 
at three time step size
of $\epsilon=0.2$ (large circles), $\epsilon=0.05$ (smaller circles)
and $\epsilon=0.01$ (dotted line). The connecting line segments are
for guiding the eye only.
}
\label{fsix}
\end{figure}
\end{document}